\documentclass[preprint,nofootinbib]{revtex4-1}
\usepackage{graphicx}
\usepackage{graphicx,array,dcolumn}
\usepackage{calc,tabularx, epsfig,mathrsfs}
\usepackage{hyperref}
\usepackage{tikz}
\usepackage{amsmath,verbatim,enumerate}
\usepackage{amssymb}
\usepackage{multirow}

\usepackage{bm}
\usetikzlibrary{patterns}
\def \o{(\mathbf{k},\mathbf{\bar{k}})}
\def \op{(\mathbf{k},\mathbf{\bar{k}}')}
\def  \al{\alpha}
\def \om{\omega}
\def \bom{\bar{\omega}}

\def \x{\mathbf{x}}
\def \omk{\mathbf{k}}
\def \bomk{\mathbf{\bar{k}}}

\def \O{\mathcal{O}}

\def \fa{f_{A,\omk}(r_A,\mathbf{x}_A)}
\def \fb{f_{B,\bomk}(r_B,\mathbf{x}_B)}
\def \fbs{f_{B,\bomk}^*(r_B,\mathbf{x}_B)}

 \def \ddom{d^d \mathbf{k}}
  \def \ddbom{d^d \mathbf{\bar{k}}}
\def\be{\begin{equation}}
\def\ee{\end{equation}}
\def \ell{\mathbf{\Lambda}}
\def\bal{\begin{align}}

\def\eal{\end{align}}
\begin{document}

\title{Symmetry transformation of subregion bulk representations }

\author{Nirmalya Kajuri}
\email{nirmalya@iitmandi.ac.in}
\affiliation{Indian Association for Cultivation of Science, Kolkata 700032 \\}
\affiliation{Indian Institute of Technology Mandi,\\ Kamand, Mandi-175005}

\begin{abstract}
Different AdS-Rindler wedges can be mapped to each other using bulk isometries. In this paper we address how the boundary representations corresponding to the AdS-Rindler wedges transform under such isometries. We show that when a bulk wedge is mapped to another using a bulk isometry, their boundary representations are mapped by the conformal transformation corresponding to the isometry. We comment on the import of this result on the relation between AdS/CFT and quantum error correction.

\end{abstract}

\maketitle

\section{Introduction.} 

Shortly after the discovery of the AdS/CFT correspondence \cite{Maldacena:1997re,Gubser:1998bc,Witten:1998qj}, it was noted \cite{Banks:1998dd} that in a large-N CFT, the algebra of creation and annihilation operators for a bulk scalar is reproduced by the following CFT operators:
\begin{align} \label{anni}
 \O_{\om,k}= \int dt \, d^{d-1}x \,e^{i \om t -ik \cdot x}\O(t,x) \\
  \O_{-\om,-k}= \int dt \, d^{d-1}x \, e^{-i \om t+ik \cdot x}\O(t,x)
\end{align}

where $\O(\x)$ is a single trace primary operator.

Indeed, the extrapolate dictionary \cite{Balasubramanian:1998sn,Banks:1998dd} can be used to relate the bulk creation and annihilation operators to the above CFT operators:

\begin{align} \label{ads}
a_{\om,k} &= \frac{1}{N_{\om,k}} \O_{\om, k}\\
a^\dagger_{\om,k} &= \frac{1}{N_{\om,k}} \O_{-\om,-k}
\end{align}

where $N_{\om,k}$ is a normalization factor. This expression is to be understood to hold within correlators.

If one now uses \eqref{ads} and substitutes the operators $O_{\om,k}$ in place of $a_{\om,k}$ in the mode expansion of a bulk field, one obtains a boundary operator that exactly reproduces the bulk field correlators. This operator is called the boundary representation of a bulk field. The boundary representation is clearly a non-local operator in the CFT, being smeared over the boundary. It is typically expressed in the following form:

\be 
\phi(X) = \int \,d^d \x\, K(X;\x) \O(\x)
\ee

where $\x$ is the boundary coordinate and $\O(\x)$ is the CFT primary dual to the bulk field $\phi(X)$, $X$ being the bulk coordinate. The function $K(X;\x)$ is called the smearing function.

Obtaining boundary representations of bulk fields is usually termed as `bulk reconstruction' and has been carried out for different bulk fields \cite{Dobrev:1998md,Bena:1999jv,Hamilton:2005ju,Hamilton:2006az,Kabat:2011rz,Kabat:2012hp,Papadodimas:2012aq,Kabat:2012av,Kabat:2013wga,Morrison:2014jha,Sarkar:2014dma,Kim:2016ipt,Kabat:2017mun,Foit:2019nsr,Kajuri:2019kmr}. For a recent review, see \cite{Kajuri:2020vxf}.

The AdS-Rindler wedge is a coordinate patch in the bulk, which covers only part of the full AdS spacetime. One can carry out the same program by defining the operators given in \eqref{anni} entirely within this wedge. For a given AdS-Rindler wedge $A$, this would give us a boundary representation:

\be 
\phi_A(X) = \int_{ \text{boundary of}\,A} \,d^d \x\, K_A(X;\x) \O(\x)
\ee

Here, the range of integration in \eqref{anni} would be over the entire conformal boundary of the AdS-Rindler wedge. It turns out that for dimensions higher than two, the smearing function $K_A$ is a distribution rather than a function\cite{Morrison:2014jha}. As any bulk causal wedge of a ball-shaped boundary region is isometric to the AdS-Rindler wedge, boundary representations can be constructed for all such wedges. In a slight abuse of notation, we will refer to all wedges isometric to an AdS-Rindler wedge as AdS-Rindler wedges. 

The question we are interested in asking in this paper is: how do the boundary subregion representations like $\phi_A$ transform under the conformal symmetries of the boundary theory? For the global representation, the answer is known\cite{Nakayama:2015mva} (see also \cite{Miyaji:2015fia,Nakayama:2016xvw,Goto:2017olq}). The transformation rule is such that consistency between boundary conformal symmetries and bulk isometries is maintained. It is given by:

\begin{equation} \label{equation}
U(\Lambda) \phi (X) U^{-1}(\Lambda) = \phi (\ell^{-1} X)
\end{equation}

where $\Lambda$ is a boundary conformal transformation while $\mathbf{\Lambda}$ is the corresponding bulk isometry. $U(\Lambda)$ is the unitary operator in the boundary theory that implements the conformal transformation $\Lambda$.

But \eqref{equation} can not hold generally for AdS-Rindler wedges since they are not preserved under all bulk isometries. In general, an AdS-Rindler wedge will be mapped to a different AdS-Rindler wedge by a bulk isometry. We will show in this paper that an analogous relation does hold -- for an isometry that maps an AdS-Rindler wedge $A$ to a different wedge $B$, the correct transformation rule (previously conjectured in \cite{Kajuri:2019kmr}) is:
\be\label{symmetry}
U \phi_A(X_A)U^{-1}= \phi_B(X_B)
\ee
where $X_B$ is the image (in chart $B$) of $X_A$ under the isometry that maps the two charts.

This is quite similar to the first transformation law, except that now one AdS-Rindler wedge gets mapped into another. This is what one would expect, given that bulk isometries which map different AdS-Rindler wedges turn into conformal symmetries in the boundary limit. This conformal symmetry is realized as a unitary map in the CFT\cite{Casini:2011kv}. It is therefore natural to expect that it is this unitary map that relates the  boundary representations in the different wedges.

We may recast \eqref{symmetry} in the language of Bogoliubov transformations. This is useful in discussions of overlapping regions. Generally, if we have different sets of creation and annihilation operators corresponding to the same field, they are related by Bogoliubov transformations: 

\be \label{annitr}
a_{\om,k}  = \sum_{\bom,\bar{k}} \al(\om,k;\bom,\bar{k} )b_{\bom} +\beta (\om,k;\bom,\bar{k}) b^\dagger_{\bom,\bar{k}}
\ee

One might ask: what are the boundary analogs of the above equations? Or in other words, how are the $\O_\om$ operators related to each other? At first glance, it appears that the two can't be related. If we consider $\O_{A,\om},\O_{B,\bom}$ as the boundary representations of annihilation operators for AdS/Rindler wedges $A,\, B$ respectively, they cannot satisfy a relation like \eqref{annitr} simply because of the fact that they have support over different boundary regions.

 It is straighforward to see that \eqref{symmetry} gives us a map between the different sets of annihilation/creation operators. This is given by:
 
\be \label{trn}
U \O^A_{\om,k} U^{-1} \sim \int d^d\bom d^d\bar{k} \, \left( \al(\om,k;\bom,\bar{k} )\O^B_{\bom,\bar{k}} + \beta(\om,k ; \bom,\bar{k} ) \O^B_{-\bom,\bar{k}} \right)
\ee

This resembles \eqref{annitr} except for the insertion of the unitaries, just like \eqref{symmetry} resembles a coordinate change in the bulk. The general lesson here is that the relations that map bulk objects in different charts is implemented by CFT unitaries on the  AdS-Rindler boundary representations of the same field.

In the next section, we will prove \eqref{symmetry}. In the third section, we will discuss consequences that follow from \eqref{symmetry}. We conclude with a summary.

\section{Transformation law for AdS/Rindler representations}

Let us consider an AdS-Rindler wedge $A$ with a coordinate chart $(r_A,\x_A)$ where $r_A$ is the radial coordinate and $x_A$ are the boundary coordinates. The boundary representation of the bulk field in this wedge is given by :
\be 
\phi_A(r_A,\x _A)= \int d^d \x _A K(r_A,\x _A;\x'_A)\O(\x'_A).
\ee
Here the integration is over the boundary of the wedge $A$.

A general isometry will map it to a different wedge $B$ whose coordinate chart is denoted as $(r_B,\x_B)$ and the corresponding boundary representation as $\phi_B(r_B,\x_B) $. This isometry will reduce to a  conformal symmetry on the boundary. The latter will be implemented by a unitary operator in the boundary CFT, which we call $U_{AB}$. 

We want to show that
\be \label{symm}
U_{AB}\, \phi_A(r_A,\x_A){(U_{AB})}^{-1}= \phi_B(r_B,\x_B)
\ee

where $(r_B,\x_B)$ is the image (in the chart $B$) of $(r_A,\x_A)$  under the isometry that maps the two wedges.

Applying this unitary on $\phi_A(r_A,\x_A)$ we get: 
\begin{align}
\notag U_{AB}\, \phi_A(r_A,\x_A){(U_{AB})}^{-1}=& \int d^d \x_A \, K_A(r_A,\x_A;\x'_A)U_{AB}\, \O(\x'_A){(U_{AB})}^{-1}\\
&\notag =\int d^d \x_A \, K_A(r_A,\x_A;\x'_A)\left|\frac{\partial \x'_B}{\partial x'_A} \right|^\Delta \O(\x'_B)\\
\notag &= \int d^d \x_B \left|\frac{\partial \x'_B}{\partial \x'_A} \right|^{\Delta -1} K(r_A,\x_A;\x'_A) \O(\x'_B)\\ \label{plymuth}
& =  \int d^d \x_B \left|\frac{\partial \x'_B}{\partial \x'_A}\right|^{\Delta -1} K'_A(r_B,\x_B;\x'_B) \O(\x'_B)
\end{align}
where $$K'_A(r_B,\x_B;\x'_B)=K_A(r_A,\x_A;\x'_A)  $$.
In the first step, we have used the unitary transformation $U_{AB}\,\O_A(\x_A){(U_{AB})}^{-1}=\left|\frac{\partial \x'_B}{\partial x'_A} \right|^\Delta \O_B(\x_B)$. In the next step, we have simply changed the integration variable. In the final step, we expressed the smearing function in the coordinates of chart $B$. Now the integration is over the boundary of the wedge $B$.

So we indeed have an operator with the same support as $\phi_B$. We now proceed to show that it is the same operator.

The smearing function $K_A(r_A,\x_A;\x'_A)$ is defined by:
\be  \label{smear}
K_A(r_A,\x_A;\x'_A) =\int \ddom \,f_{A,\omk}(r_A,\x_A)\, e^{-i \omk \cdot \x_A} 
\ee 

where $\omk = (-\om,k)$. Here, $f_{A,\omk}(r_A,\x_A)$ is a normalizable mode solution of the Klein-Gordon equation in the coordinate chart $A$. The solution is not normalized. To normalize it one has to multiply it with the normalization factor, which we denote as $N_{\omk}$.

We also have that \cite{Hamilton:2006az,Morrison:2014jha}
\be \label{gunnarsson}
 e^{i \omk \cdot \x_A} = \lim_{r_A \to \infty} r^\Delta f_{A,\omk}(r_A,\x_A)
\ee

As the two wedges are related by a coordinate transformation, we can express the modes $\fa =f'_{A,\omk}(r_B,\x_B)  $ as a linear combination of the modes $\fb$:

\be 
 N_{\omk} f_{A,\omk}(r_A,\x_A)=f'_{A,\omk}(r_B,\x_B)=\int \ddbom\, \left( \al\o (M_{\bomk}\,  \fb) + \beta \o (M_{\bomk}\,  \fbs)  \right)
 \ee
where $N_{\omk}, M_{\bomk}$ are the normalization constants for the mode solution $\fa, \fb$ respectively. We will also have an analogous relation to \eqref{gunnarsson}:
\be \label{gunnarssonson}
 e^{i \bomk \cdot \x_B} = \lim_{r_B \to \infty} r^\Delta f_{B,\bomk}(r_B,\x_B)
\ee

For this transformation to be a canonical transformation in the bulk field theory the following well-known conditions must hold:
\be \label{tight}
\int \ddom\, \left(\al\o \al^*\op - \beta \o \beta^* \op \right)  =\delta(\bomk-\bomk').
\ee
\be \label{light}
\int \ddom\, \left(\al\o \beta \op - \beta \o \al \op \right)  =\delta(\bomk-\bomk').
\ee

Using \eqref{gunnarsson} and\eqref{gunnarssonson}, we can deduce how $ e^{i \omk \cdot \x_A} $ and $ e^{i \bomk \cdot \x_B} $ are related:

\begin{align} 
\notag \lim_{r_A \to \infty} {r_A}^\Delta \fa &=  \lim_{r_B \to \infty}{r_A}^\Delta\frac{1}{N_{\omk}}\int \ddbom \, M_{\bomk}\left(\al\o \fb+ \beta \o \fbs\right)\\ \label{barely}
\implies e^{i \omk \cdot \x_A}    &=  \left|\frac{\partial \x'_B}{\partial \x'_A}\right|^{\Delta }\frac{1}{N_{\omk}} \int \ddbom \, M_{\bomk}\, \left(\al\o e^{i \bomk \cdot \x_B} +\beta \o e^{-i \bomk \cdot \x_B}\right)
\end{align}

Here we have used $r_A =r_B \left|\frac{\partial x'_B}{\partial x'_A} \right| $ (See for instance\cite{Belin:2018juv}). 

From this, one can deduce a similar relation for $e^{-i \omk \cdot \x_A}$ with a little calculation. This is given by:

\be \label{rohit}
e^{-i \omk \cdot \x_A}=\left|\frac{\partial \x'_B}{\partial \x'_A}\right|^{1-\Delta }N_{\omk} \int \ddbom \, \frac{1}{M_{\bomk}}\, \left(\al^*\o e^{i \bomk \cdot \x_B} -\beta^* \o e^{-i \bomk \cdot \x_B}\right)
\ee

An easy way to check \eqref{rohit} is to multiply the LHS with that of \eqref{barely} (for two different coordinates $\x_A$ and ${\x_A}'$) and similarly for the RHS, then integrate over $\omk$. One can then check using \eqref{tight} and \eqref{light} that both sides will give the same delta function. 

Using \eqref{rohit} in \eqref{smear} and using \eqref{tight},\eqref{light} we finally get:
\begin{align}
\notag K_A(r_A,\x_A;\x'_A) &=K'_A(r_B,\x_B;\x'_B) = \int \ddom \, f_{A,\omk}(r_A,\x_A)e^{-i \omk \cdot \x_A} \\ \notag &=\left|\frac{\partial \x'_B}{\partial \x'_A}\right|^{1-\Delta } \int \ddbom \, \fb e^{-i \bomk \cdot \x_B} \\ \label{bristol} &=\left|\frac{\partial \x'_B}{\partial \x'_A}\right|^{1-\Delta }K_B(r_B,\x_B;\x'_B)
\end{align}

Substituting \eqref{bristol} in \eqref{plymuth}, we obtain \eqref{symm}. We have thus proven \eqref{symm}. 

Now we consider the boundary representations of the AdS-Rindler creation and annihilation operators. For the wedge $A$, they are defined by: 
\be 
{\O^A}_\omk = \int \, d^d \x_A \, \O(\x_A) e^{-i \omk \cdot \x_A}
\ee

A similar definition holds for ${ \O^B }_{\bomk}$.

Following the same steps as the proof above, it can be shown immediately that: 
\be \label{bog}
U_{AB} {\O^A}_\omk {(U_{AB})}^{-1}=N_{A,\omk} \int \ddbom \, \frac{1}{M_{B,\bomk}} \left( \al^*\o { \O^B }_{\bomk} - \beta^* \o  { \O^B }_{-\bomk} \right)
\ee

Which is what we expected. So we see that analogs of Bogoliuboov transformations do hold in the boundary. 

We note that \eqref{bog} and \eqref{symm} are exact operator relations.

\section{ Implications and Examples}  

Certain paradoxes arise when overlapping wedges are considered \cite{Almheiri:2014lwa}. Our result may help shed some light on some of these paradoxes.

 Let us consider two overlapping AdS-Rindler wedges $A$ and $B$. The  corresponding boundary representations  of a scalar field are $\phi_A$ and $\phi_B$. By construction, we should have that:
 
 \be \label{prdss}
\langle \phi_A(X)...\phi_A(X') \rangle =\langle \phi_B(X)...\phi_B(X') \rangle
\ee

This gives rise to a paradox, because $\phi_A(X)$ and $\phi_B(X)$ are manifestly two different operators in the CFT, and all their correlators cannot agree. Further paradoxes arise when one asks, where This paradox may be resolved by demanding that the equality holds only in a subspace of the CFT Hilbert space, which is called the code subspace \cite{Almheiri:2014lwa}. There are further paradoxes which have to do with the question of where in the boundary is the information about a bulk region located, which we will not discuss here (in \cite{Almheiri:2014lwa}, it was suggested that these can be resolved using quantum error correction). 

However, our results may shed some light on the equality \eqref{prdss}. Let us consider two overlapping wedges $A$ and $B$. We know that $A$ can be mapped to $B$ via some isometry $\Lambda$. Then it follows from \eqref{symm} that: 

\be \label{prdx}
\langle \phi_A(X_A)\phi_A(X'_A) \rangle =\langle \phi_B(X_B)\phi_B(X'_B) \rangle
\ee

Now the invariance of correlation functions under isometry for a given AdS-Rindler representation holds by construction:
\be \label{prd}
\langle \phi_B(X)\phi_B(X') \rangle =\langle \phi_B(\Lambda^{-1}X)\phi_B(\Lambda^{-1}X') \rangle
\ee

Then we have that:

\be \label{prds}
\langle \phi_A(X)\phi_A(X') \rangle =\langle \phi_B(X)\phi_B(X') \rangle
\ee

Thus we see that we may be able to derive \eqref{prdss} directly. 

\begin{figure}
\centering

 \includegraphics[width=.7\linewidth]{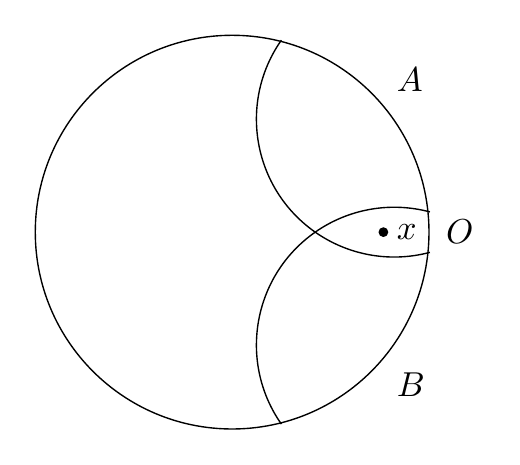}

\caption{The figure shows a top view of two overlapping AdS/Rindler wedges $A$ and $B$. $O$ is the overlap region between the two. The two wedges can be mapped to each other via an isometric map. Here we assume the case where the points in $O$ get mapped to themselves under this map. So a point in the overlap $x$ will have a co-ordinate $X_A$ in the chart $A$ and $X_B$ in the chart $B$, and $X_A$ will be the image of $X_B$(and vice versa) under this map.}
\label{fig}
\end{figure}

When the overlapping wedges are related by rotations, the points in the overlapping region simply gets mapped to themselves. In this special case, \eqref{prds} follows directly from \eqref{symm}. This is depcited in \ref{fig}.

\begin{figure}
\centering

 \includegraphics[width=.7\linewidth]{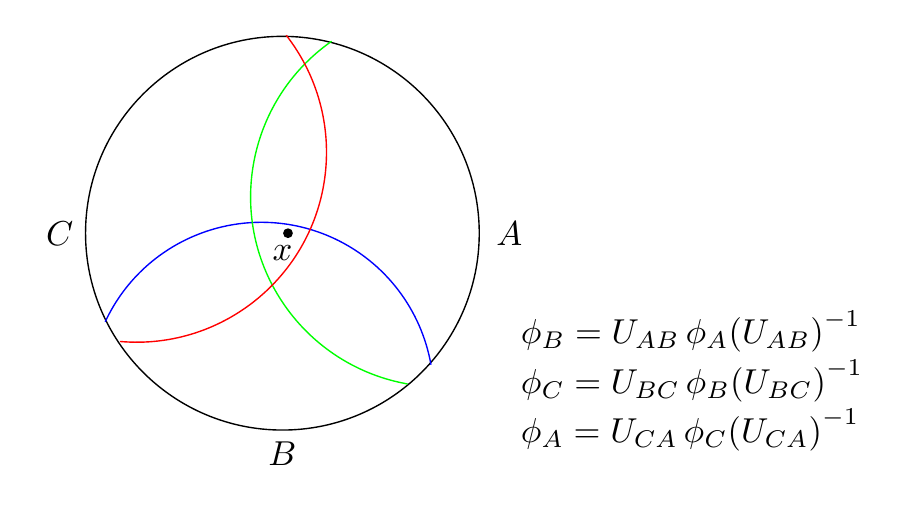}

\caption{The figure shows a top view of three overlapping AdS/Rindler wedges $A$ ,$B$ and $C$. $x$ is a point in the overlap region between the three. The boundary representations corresponding to the three wedges are related by unitary operators.}
\label{picapica}
\end{figure}

So we see that we can prove the equivalence of the correlators in the overlapping regions of  AdS-Rindler wedges from the operator relation\eqref{symm}. This may help shed some light on how bulk information is stored in overlapping wedges. Indeed, in the special case of wedges related by rotational symmetry in the bulk, it appears that the different subregion boundary representations can be mapped to each other via unitary maps (This is depicted in figure \ref{picapica}). Thus for AdS-Rindler wedges,\eqref{symmetry} could be a possible explanation of why different \eqref{prds} holds in the code subspace.

We now note another consequence of  \eqref{symm}. If we consider a cross-correlator between two subregion representations,  \eqref{symm} dictates that we will obtain:

\be \label{huh}
\langle \phi_A(X)\phi_B(X') \rangle  = \langle \phi_A(X)U_{AB}\phi_A(X') \rangle 
\ee

Naively, one would have expected that the LHS would equal $\langle \phi_A(X)\phi_A(X') \rangle $. But this cannot be the case. For bulk fields, the relation $\langle \phi_A(X)\phi_B(X') \rangle = \phi_A(X)\phi_A(X') \rangle $ holds because bulk isometries map the field to itself. However, this corresponds to a unitary map acting on the boundary representation of the bulk field. In other words, \eqref{huh} is a consequence of the fact that the boundary representations of creation and annihilation operators corresponding to different wedges are not linear combinations of each other, unlike their bulk counterparts. Instead, \eqref{bog} is how they map to each other.

 Hence, one cannot replace bulk fields by their boundary representations in a cross-correlator. This fact can also be seen from considering the extrapolate dictionary for the LHS of \eqref{huh}. In the boundary limit, the term in the LHS will go to a CFT correlator $\langle\O(\x_A)\O({x_B}')\rangle$. As the two primaries are on two different charts on the boundary, one has to perform a conformal transformation via a unitary operator on one of them to bring them to the same chart. This will give us a correlator of the form $\langle\O(\x_A) U\O({\x_A}')\rangle$, which is what we expect from the arguments above.

\section{Conclusions}

In this paper, we investigated the interplay between bulk isometries and boundary conformal symmetries for AdS-Rindler representations. Specifically, we asked: if we take two wedges which can be related via a bulk isometry, how are the corresponding boundary representations of bulk fields related? The answer, as we proved, is that they will be related via a conformal symmetry. 

This was, of course, the expected result. Very generally, any coordinate transformation in the bulk maps to a conformal transformation in the boundary, the action of an isometric change of coordinate charts on a bulk field is implemented by a unitary transformation on its boundary representation. 

We also discussed the implication of this result for overlapping wedges and argued that it can help shed light on how bulk information is stored in boundary code subspace.

In this paper, we have related boundary representations for two AdS/Rindler wedges via unitary transformations. However, our proof can be straightforwardly extended to the general case and can be used to relate boundary representations on different coordinate charts (such as global and AdS-Rindler charts, for instance). In a parallel work\cite{Dey:2021vke}, we have shown that for the different charts in $AdS_2$, the boundary representations are indeed related by conformal symmetries. We plan to consider higher dimensional cases in the future.

\end{document}